\documentclass[12pt,a4paper,dvips]{article}
\usepackage{a4p}
\usepackage{cite,mcite}
\usepackage{graphicx}
\usepackage{physics }
\usepackage{l3_title,ifthen}
\usepackage{rotating}
\journalname{Phys. Lett. B}

\usepackage{Lep}
\Lep{2}

\preprint{2000-006}
\date{January 11, 2000}

\newlength{\capindent}
\newlength{\capwidth}
\newlength{\figwidth}
\newcommand{\icaption}[2][!*!,!]{\hspace*{\capindent}%
  \begin{minipage}{\capwidth}
    \ifthenelse{\equal{#1}{!*!,!}}%
      {\caption{#2}}%
      {\caption[#1]{#2}}
  \end{minipage}}
\begin{document}
\setlength{\unitlength}{1mm}
\begin{titlepage}
\title{Measurement of the \boldmath{${\epem\ra\Zo\gamma\gamma}$}
Cross Section and Determination of Quartic Gauge Boson Couplings at LEP}
\author{The L3 Collaboration}
\begin{abstract}

A first measurement of the cross section of the process
$\epem\ra\Zo\gamma\gamma$  is reported using a total
 integrated luminosity  of 231\,pb$^{-1}$
collected with the L3 detector 
at centre-of-mass energies
of $182.7\GeV$ and $188.7\GeV$. By selecting
hadronic events with two isolated photons
the $\epem\ra\Zo\gamma\gamma$ cross section is measured to be
$ 0.49^{+0.20}_{-0.17}\pm
0.04\,{\mathrm{pb}}$ at $182.7\GeV$ and  $0.47 \pm 0.10\pm
0.04\,{\mathrm{pb}}$ at $188.7\GeV$. The
measurements are consistent  with Standard Model expectations.
Limits on  Quartic Gauge Boson Couplings 
$a_0/\Lambda^2$ and $a_c/\Lambda^2$ of
$-0.009\GeV^{-2}  < a_0/\Lambda^2 < 0.008\GeV^{-2}$ and
$-0.007\GeV^{-2}  < a_c/\Lambda^2 < 0.013\GeV^{-2}$ are derived at
95\% confidence level. 

\end{abstract}
\submitted

\end{titlepage}

%
%%%%%%%%%%%%%%%%%%%%%%%%%%%%%%%%%%%%%%%%%%%%%%%%%%%%%%%%%%%%%%%%%%%%%%%%%%%%%%%
% Introduction
%%%%%%%%%%%%%%%%%%%%%%%%%%%%%%%%%%%%%%%%%%%%%%%%%%%%%%%%%%%%%%%%%%%%%%%%%%%%%%%
%

\section{Introduction}

The LEP centre-of-mass  energy ($\sqrt{s}$) for $\epem$ collisions
has now exceeded the  W pair
and Z pair production thresholds, allowing the study of triple gauge
boson production processes such as $\epem\ra\Zo\gamma\gamma$ and
$\epem\ra\Wp\Wm\gamma$. Measurements of these processes give a new
insight into the Standard Model of electroweak
interactions (SM)~\cite{sm_glashow}.
The possibility of these triple gauge boson production processes
proceeding via $s$-channel exchange of a fourth boson provides a probe
of quartic  gauge boson
couplings (QGC). Such measurements were  recently performed for the
$\epem\ra\Wp\Wm\gamma$ process~\cite{opalwwg}. 

\begin{figure}[p]
  \begin{center}
      \mbox{\includegraphics[width=\figwidth]{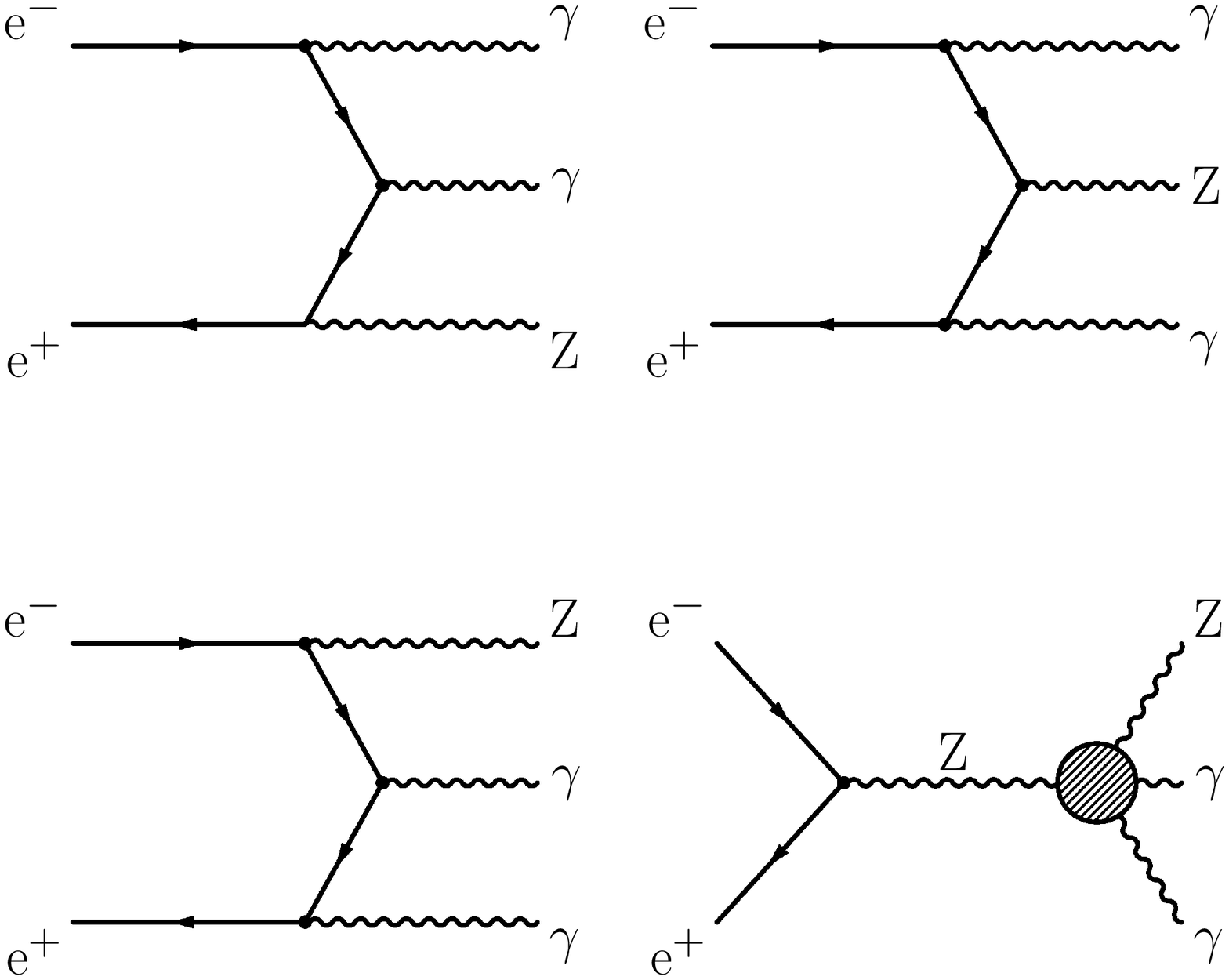}}
      \icaption{Three of the six SM diagrams contributing to
      $\epem\ra\Zo\gamma\gamma$ 
        production. The other three SM diagrams are obtained by
      crossing the photon lines. A  possible anomalous QGC diagram 
      is also shown. 
    \label{fig:0}}
  \end{center}
\end{figure}

This letter describes the first measurement of the cross section of the
process  $\epem\ra\Zo\gamma\gamma$  
followed by the hadronic decay of the Z. 
In the SM this process occurs by radiation of 
the photons from the incoming electron and/or positron, corresponding to a
total of six diagrams, three of which are presented in
Figure~\ref{fig:0}. 
No QGC contribution is predicted at the tree
level and  Z$\gamma\gamma$ events are sensitive to anomalous
QGC~\cite{bb1,sw,bb2} contributions, as shown in Figure~\ref{fig:0}.

The measurement uses
data  collected with the L3
detector~\cite{l3_00} at
LEP in 1997 and 1998 at average centre-of-mass energies of
$\sqrt{s}=182.7\GeV$
and $\sqrt{s}=188.7\GeV$ corresponding to  integrated luminosities of
55\,pb$^{-1}$ and  176\,pb$^{-1}$, respectively. These energies are
respectively denoted as $183\GeV$ and $189\GeV$ hereafter. 

The $\epem\ra\Zo\gamma\gamma\ra\qqbar\gamma\gamma$ signal is defined by three
phase space cuts: photon energies above $5\GeV$, photon angles with respect to the beam axis between
$14^\circ$ and $166^\circ$ and invariant mass of the 
primary produced quarks 
within a $\pm2\Gamma_{\rm Z}$ window around the Z mass, $\Gamma_{\rm Z}$
being the Z width. The KK2f  Monte Carlo (MC) program~\cite{KK2f}
predicts signal 
cross sections of about $0.4\,{\rm pb}$ at both energies.

%
%%%%%%%%%%%%%%%%%%%%%%%%%%%%%%%%%%%%%%%%%%%%%%%%%%%%%%%%%%%%%%%%%%%%%%%%%%%%%%%
% Event Selection
%%%%%%%%%%%%%%%%%%%%%%%%%%%%%%%%%%%%%%%%%%%%%%%%%%%%%%%%%%%%%%%%%%%%%%%%%%%%%%%
%
\section{Event Selection}

The selection of events satisfying the signal definition given above is
optimised using hadronic events generated at $\sqrt{s} = 189\GeV$
with the KK2f MC program and at  $\sqrt{s} = 183\GeV$ with the
PYTHIA\,5.72~\cite{pythia} MC program. Events from these MC programs
failing the signal definition are termed  QCD$\gamma\gamma$ background.
Other  background processes are 
generated both at  $\sqrt{s} = 183\GeV$ and
$\sqrt{s} = 189\GeV$ with the MC programs PYTHIA
($\rm e^+ e^- \rightarrow Z \epem$ and $\rm e^+ e^- 
\rightarrow ZZ$), 
KORALZ\,4.02~\cite{koralz} ($\rm e^+ e^- \rightarrow \tau^+ \tau^-
(\gamma)$), PHOJET\,1.05c~\cite{phojet} ($\rm e^+ e^- \rightarrow e^+ e^-
\qqbar$) and  KORALW\,1.21~\cite{koralw} for  $\Wp\Wm$ production
except for the $\rm e\nu_\e q\bar q'$ final states which are generated
with EXCALIBUR~\cite{exca}. Additional background sources are found to
be negligible. 

The L3 detector response is simulated using the GEANT 3.15
program~\cite{geant}, which takes into account the effects of energy loss,
multiple scattering and showering in the detector. 
Time dependent detector inefficiencies, as measured in each data taking
period, are reproduced in these simulations.

The selection of $\epem\ra\Zo\gamma\gamma\ra\qqbar\gamma\gamma$
 candidates
 from balanced
hadronic events with two photons and little 
energy deposition at low polar angles is based on  photon
energies and angles together with the invariant mass
of the hadronic system. The photon energy and angle criteria follow
 directly from the signal definition whereas the invariant mass of the
 hadronic system is required  to be  between $74\GeV$ and $116\GeV$.
 The main
background after these selection requirements is  due to the radiation
of two initial state
photons with a hadronic system failing the Z signal definition
criteria. The boost $\beta_{\rm Z}$ of the recoiling system to the
photons, assuming 
its mass to be the nominal Z mass, is on average larger for these background
events. Candidate events are hence required to have $\beta_{\rm Z}<0.64$ at 
$\sqrt{s} = 183\GeV$ and $\beta_{\rm Z}<0.66$ at $\sqrt{s} = 189\GeV$. Another class of
background events is the so called radiative return to the Z, where a
photon in the initial state is emitted, bringing the effective
$\sqrt{s}$ to the Z resonance. A Z boson is then produced
decaying into a hadronic system with an electromagnetic energy
deposition (photon, misidentified electron or unresolved $\pi^0$)
faking the least energetic photon of the signal selection. These
events are rejected by  an upper bound on the energy
$E^\gamma_1$ of the most
energetic photon
and a lower bound on the angle $\omega$ between the least energetic photon and
its closest jet. Numerically $E^\gamma_1<67.6\GeV$ at 
$\sqrt{s} = 183\GeV$ and  $E^\gamma_1<70.7\GeV$ at $\sqrt{s} =
 189\GeV$ with $\omega>17^\circ$ at both the energies.

\begin{figure}[p]
  \begin{center}
    \begin{tabular}{cc}
      \mbox{\includegraphics[width=.5\figwidth]{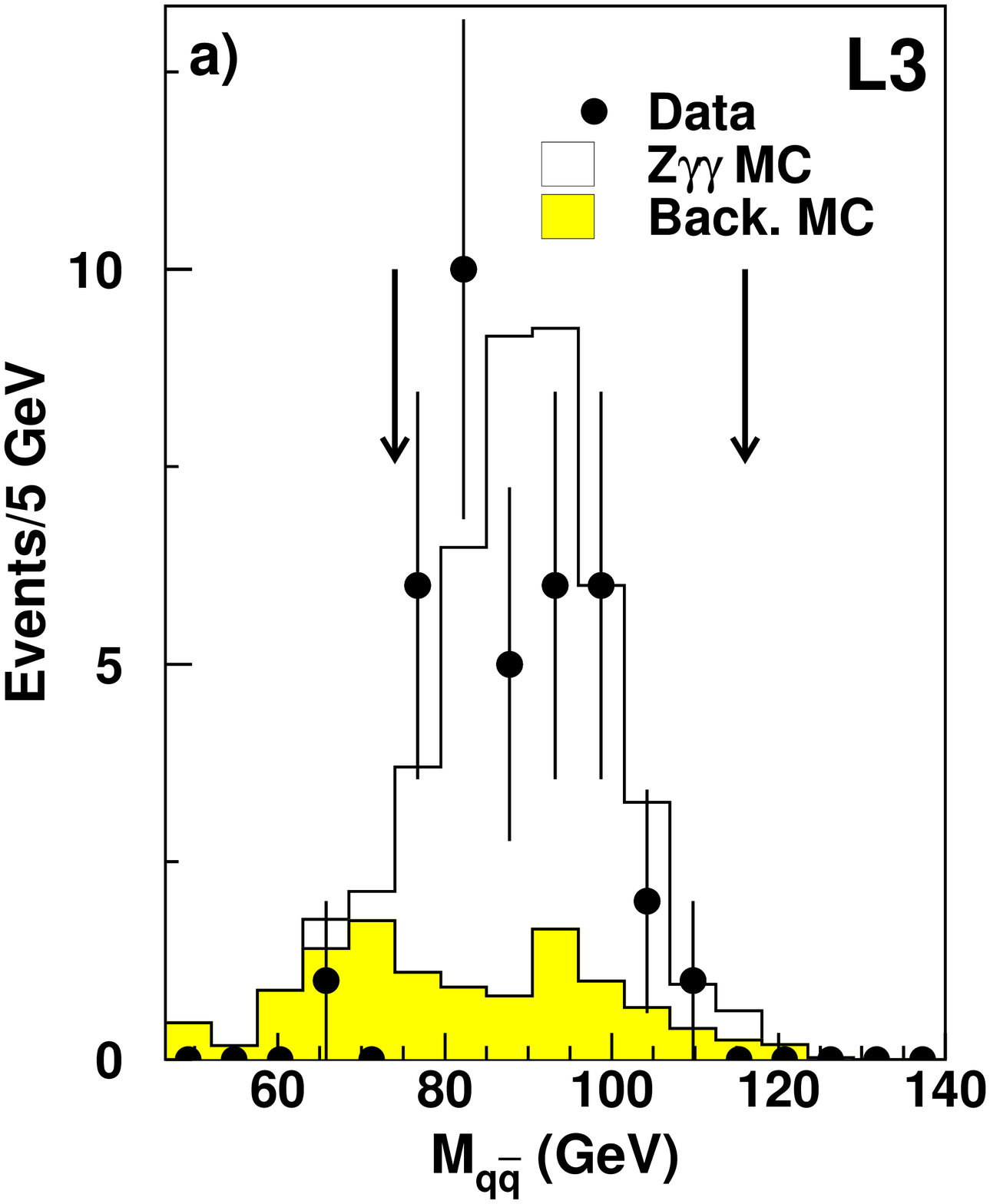}} &
      \mbox{\includegraphics[width=.5\figwidth]{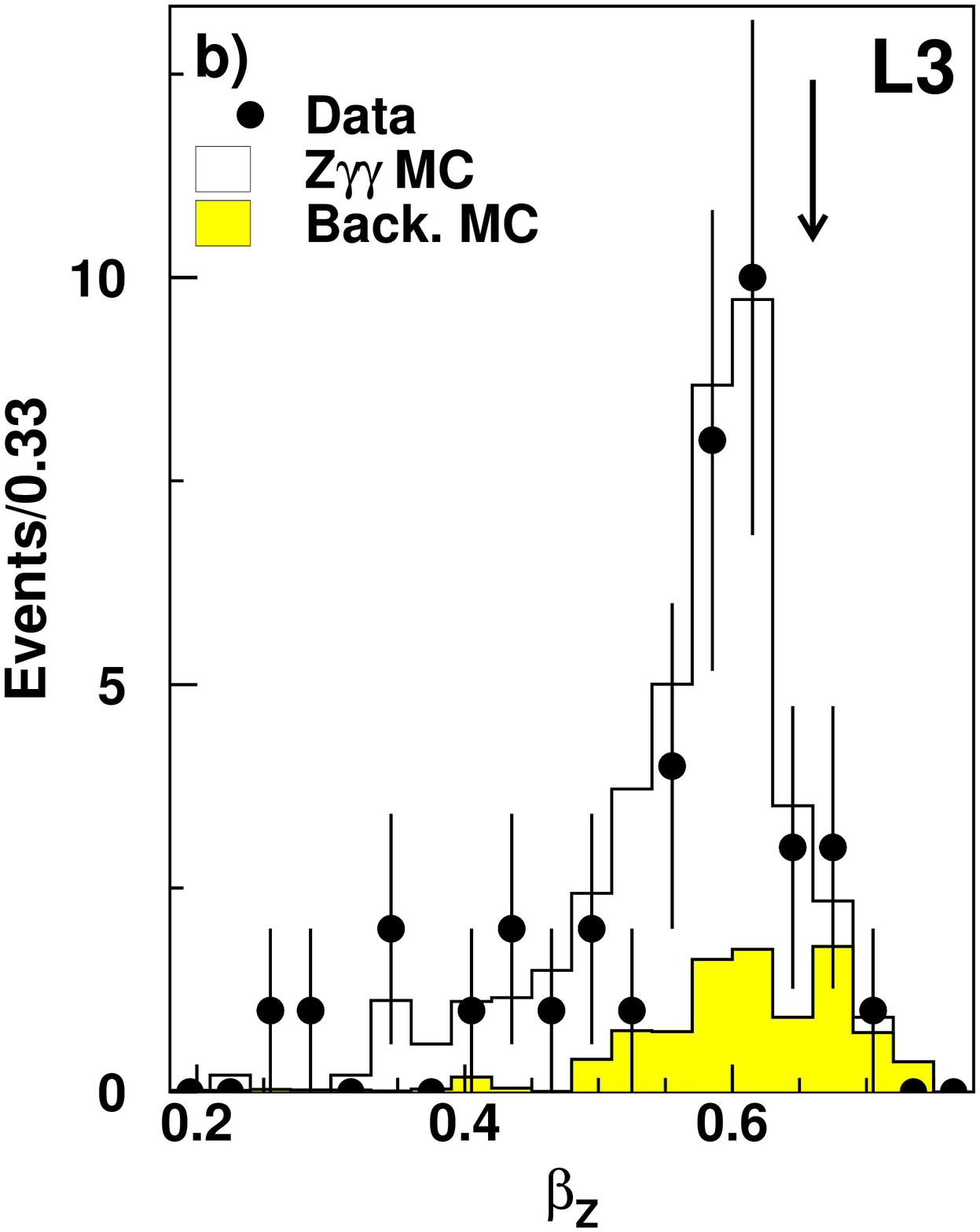}} \\
      \mbox{\includegraphics[width=.5\figwidth]{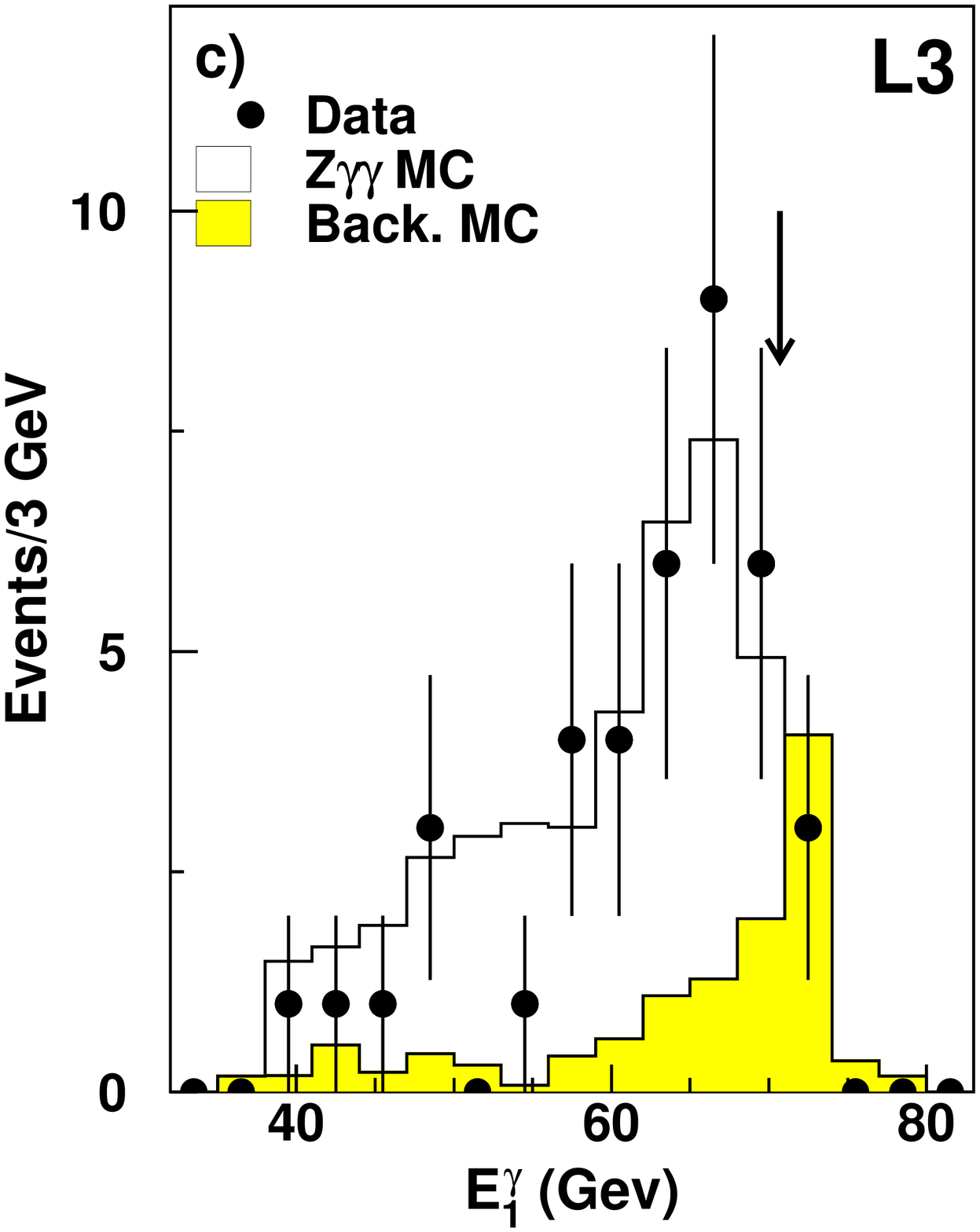}} &
      \mbox{\includegraphics[width=.5\figwidth]{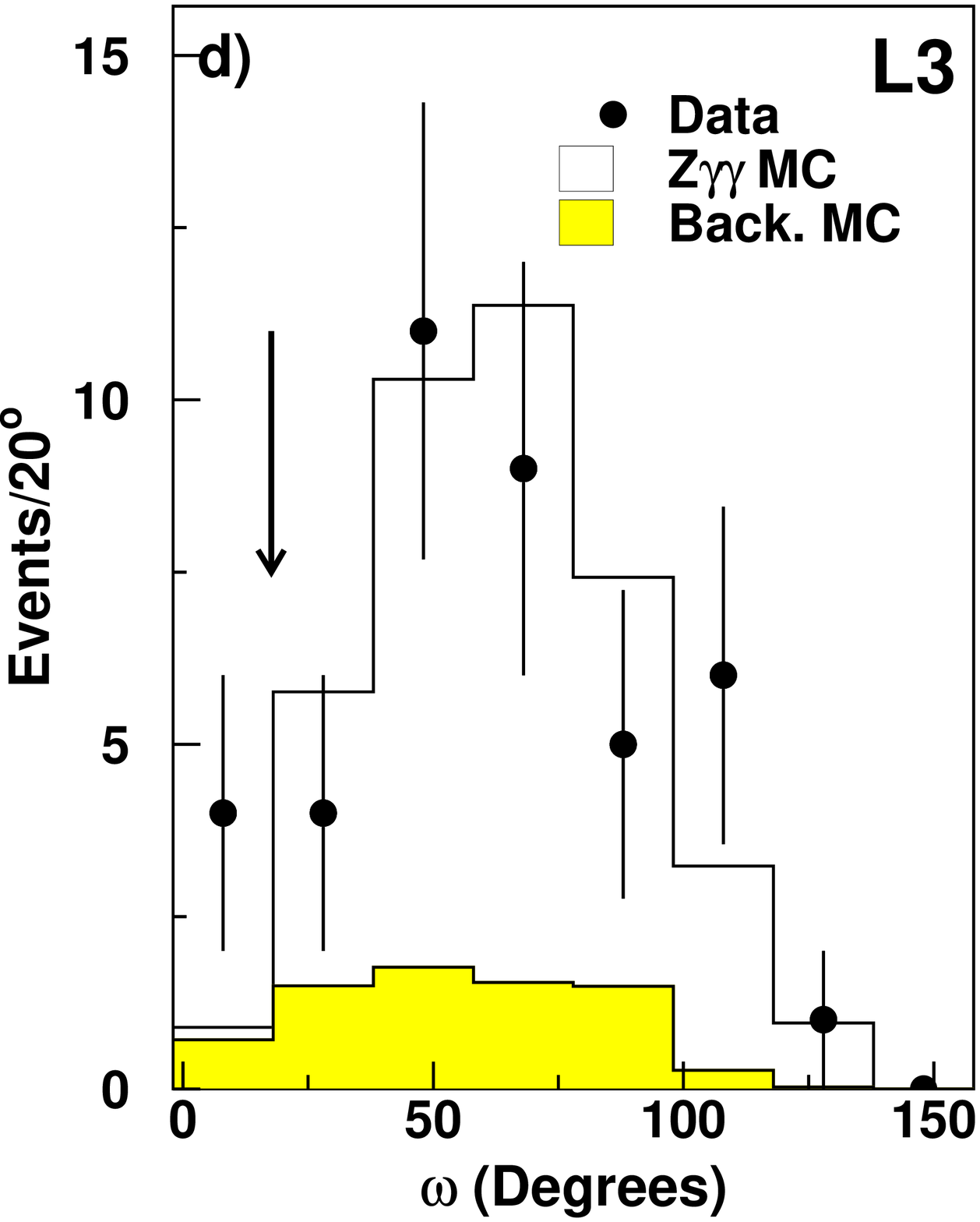}} \\
    \end{tabular}
    \icaption{Distributions of (a) the  invariant mass $M_{\qqbar}$ of
    the hadronic system, (b)
    the boost  $\beta_{\Zo}$ of the reconstructed Z boson, (c) the
    energy $E_1^\gamma$ of the most
    energetic photon  and (d) the angle $\omega$ between the least energetic
    photon and the nearest jet. Data, $\Zo\gamma\gamma$ and background MC
    are displayed for $\sqrt{s} = 189\GeV$. The
    arrows show the position of the final selection requirements. All
    the other selection criteria are applied for each plot.
    \label{fig:1}}
  \end{center}
\end{figure}

Data and MC 
distributions of the selection variables are presented in
Figure~\ref{fig:1} 
for  $\sqrt{s} = 189\GeV$, where selection
criteria on all the other variables 
are applied. A good agreement between data and MC
is observed.

%
%%%%%%%%%%%%%%%%%%%%%%%%%%%%%%%%%%%%%%%%%%%%%%%%%%%%%%%%%%%%%%%%%%%%%%%%%%%%%%%
% Results and Systematic Uncertainties
%%%%%%%%%%%%%%%%%%%%%%%%%%%%%%%%%%%%%%%%%%%%%%%%%%%%%%%%%%%%%%%%%%%%%%%%%%%%%%%
%
\section{Results and Systematic Uncertainties}

\begin{figure}[p]
  \begin{center}
    \begin{tabular}{c}
      \mbox{\rotatebox{90}{
          \includegraphics[width=.75\figwidth]{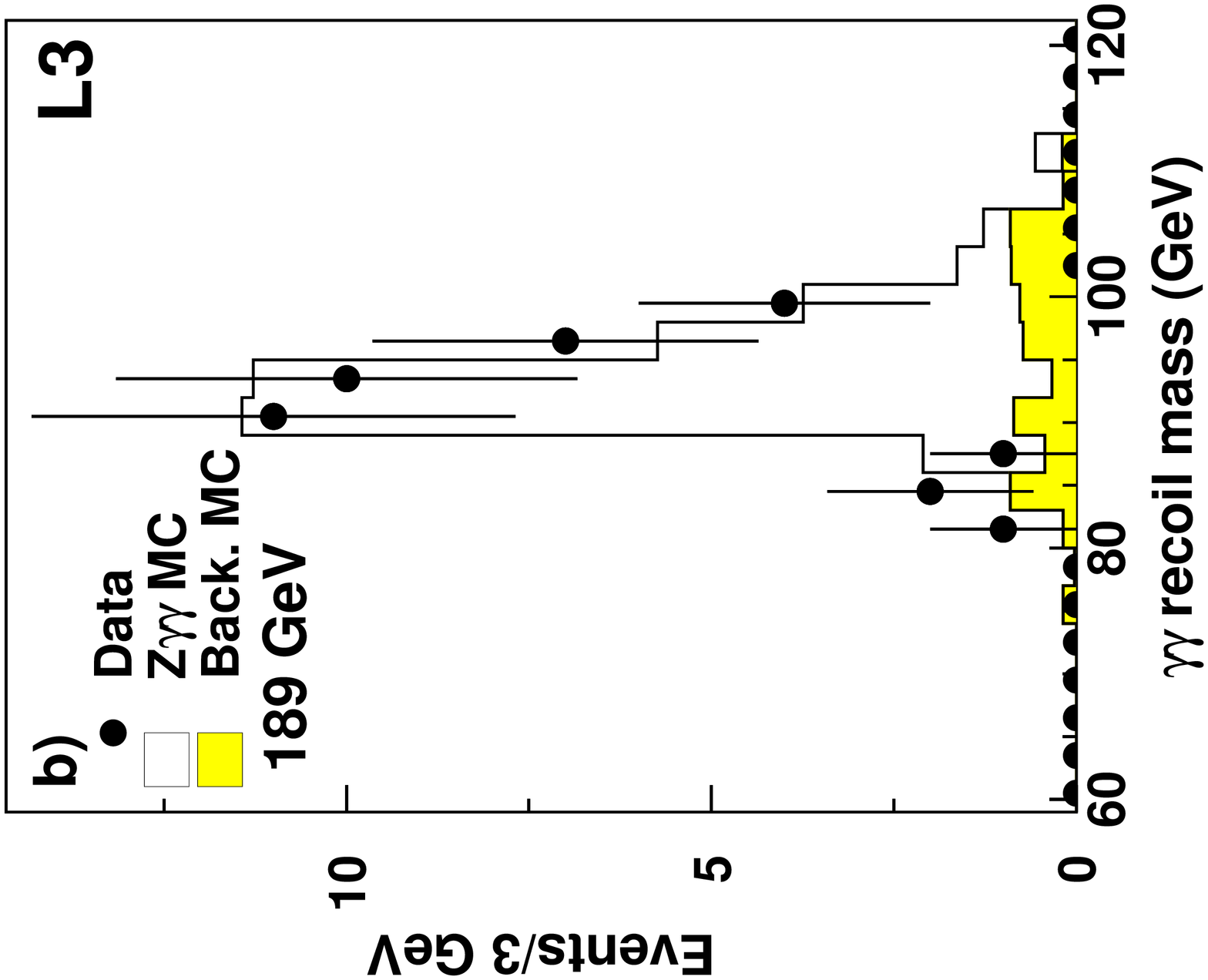}}} \\
      \mbox{\rotatebox{90}{
          \includegraphics[width=.75\figwidth]{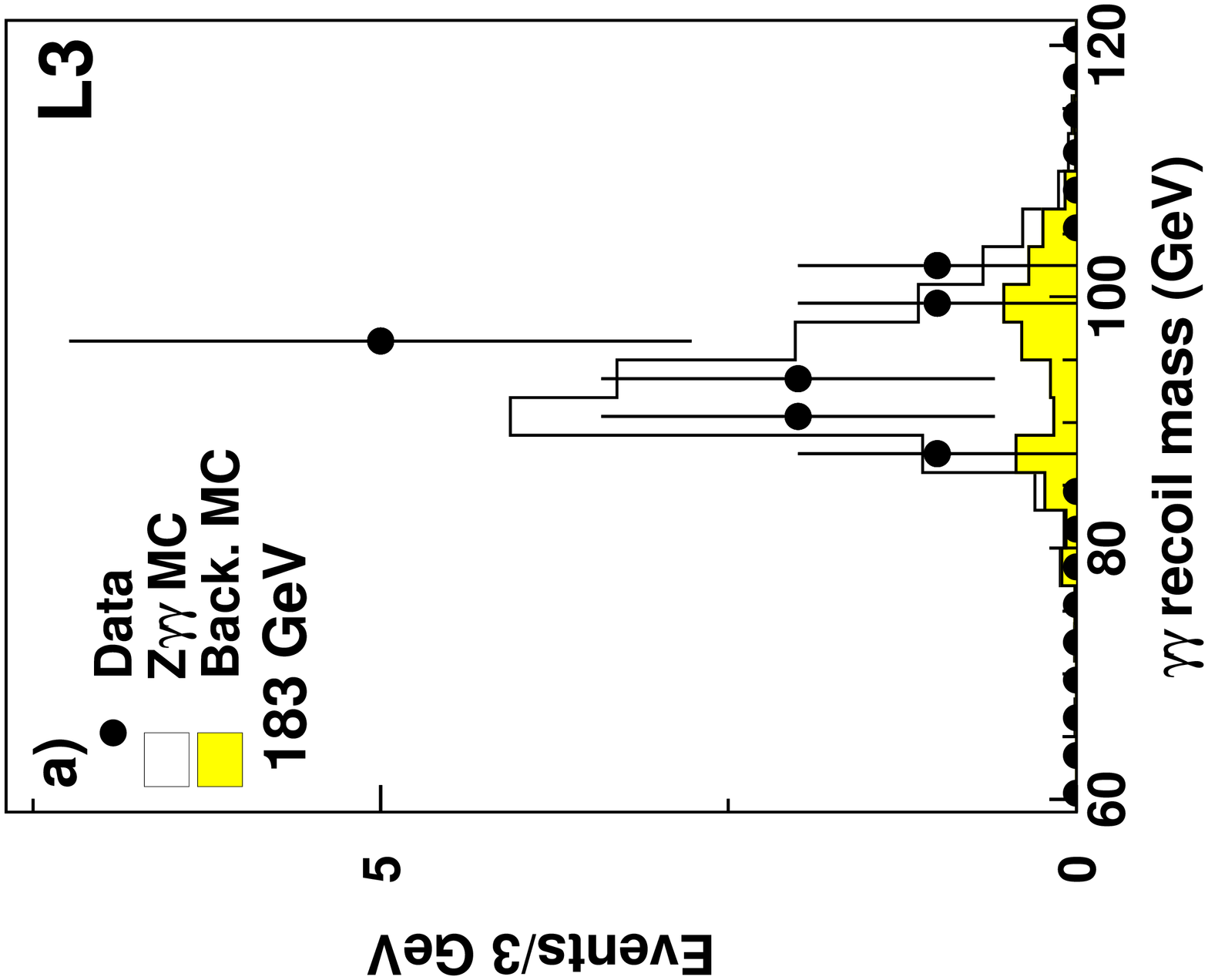}}} \\
    \end{tabular}
    \icaption{Recoil mass to the photon pairs at (a) $\sqrt{s} =183\GeV$ and
    (b) $\sqrt{s} =189\GeV$ in data, $\Zo\gamma\gamma$ and background MC.
    \label{fig:2}}
  \end{center}
\end{figure}

The application of the selection procedure described above yields the
signal efficiencies and selected data and MC events
summarised in
Table~1. A clear signal structure is observed in the recoil mass
spectra of the two photons, 
presented in Figure~\ref{fig:2} for the two  centre-of-mass energies
under study.
The $\epem\ra\Zo\gamma\gamma\ra\qqbar\gamma\gamma$ cross
section at $\sqrt{s} = 183\GeV$ and $\sqrt{s} = 189\GeV$ is then
determined from the number of events selected and the efficiency and
background estimates from MC.

\begin{table}[ht]
  \begin{center}
    \begin{tabular}{|l||r|r|}
       \hline
        $\sqrt{s}$ (GeV) &  $183$ & $189$ \\
       \hline
       $\varepsilon $          &  0.49   & 0.51    \\
       $\rm Data$                &  12   & 36   \\
       $\rm MC$                  &  13.4 & 39.2 \\
       \hline                                                       
       $\rm Z\gamma\gamma$       &  10.6 & 32.6 \\    
       $\rm QCD\gamma\gamma$     &  2.7 &   6.0 \\    
       Other                       &  0.1 &   0.6 \\    
       \hline
    \end{tabular}
    \icaption[tab:1]{Yields of the
      $\epem\ra\Zo\gamma\gamma\ra\qqbar\gamma\gamma$ selection
      at $\sqrt{s} = 183\GeV$ and $\sqrt{s} = 189\GeV$. The signal
      efficiencies $\varepsilon$ and the number of expected events for
      data and MC are 
    given.}  
  \end{center}
\end{table}

Systematic uncertainties on the cross section measurements are listed
in Table~2. They include uncertainties  arising from the signal MC
statistical error of 3\% at 
$\sqrt{s} = 183\GeV$ and  5\% at $\sqrt{s} = 189\GeV$. The uncertainty
on the accepted background
due to MC statistics are  7\% and  17\% at $\sqrt{s} = 183\GeV$ and
$\sqrt{s} = 189\GeV$, respectively.
The calorimeter energy scale uncertainty is estimated by varying
electromagnetic energies by $\pm 1\%$ and hadronic energies by $\pm
2\%$. Selection procedure systematics are obtained from the effect of
removing each of the selection criteria.  
From a comparison of the  KK2f and PYTHIA
cross sections for  hadronic events with a hard
photon, a $\pm 15\%$ uncertainty on the 
background normalisation is conservatively
estimated and  propagated to the measured cross
section. Uncertainties on signal efficiencies are estimated by comparing
KK2f with PYTHIA and GRACE~\cite{grace} MC
program predictions and are found to be negligible.
The systematic error assigned to the reweighting procedure at 
$\sqrt{s} = 183\GeV$ is estimated by applying the procedure to a
sample generated with  PYTHIA at $\sqrt{s} = 189\GeV$ and comparing
the corresponding cross section result to the previous one.

\begin{table}[ht]
  \begin{center}
    \begin{tabular}{|l||r|r|}
       \hline
       Source of  Systematics& $\Delta\sigma$ (pb)& $\Delta\sigma$  (pb) \\
       &   $183\GeV$& $189\GeV$\\
       \hline
       MC Statistics            &$ 0.02$  & $ 0.02$ \\
       Energy scale             &$ 0.02$  & $ 0.02$ \\
       Selection procedure      &$ 0.01$  & $ 0.01$ \\
       Background normalisation &$ 0.01$  & $<0.01$ \\
       Reweighting procedure    &$ 0.01$  & - \\
       \hline
       Total                    &$0.03$  & $0.03$  \\
       \hline
    \end{tabular}
    \icaption[tab:2]{Systematic uncertainties
      $\Delta\sigma$ on the
      $\epem\ra\Zo\gamma\gamma\ra\qqbar\gamma\gamma$ cross 
      section at $\sqrt{s} = 183\GeV$ and $\sqrt{s} =
      189\GeV$.}  
  \end{center}
\end{table}

\begin{figure}[p]
  \begin{center}
      \mbox{\includegraphics[width=\figwidth]{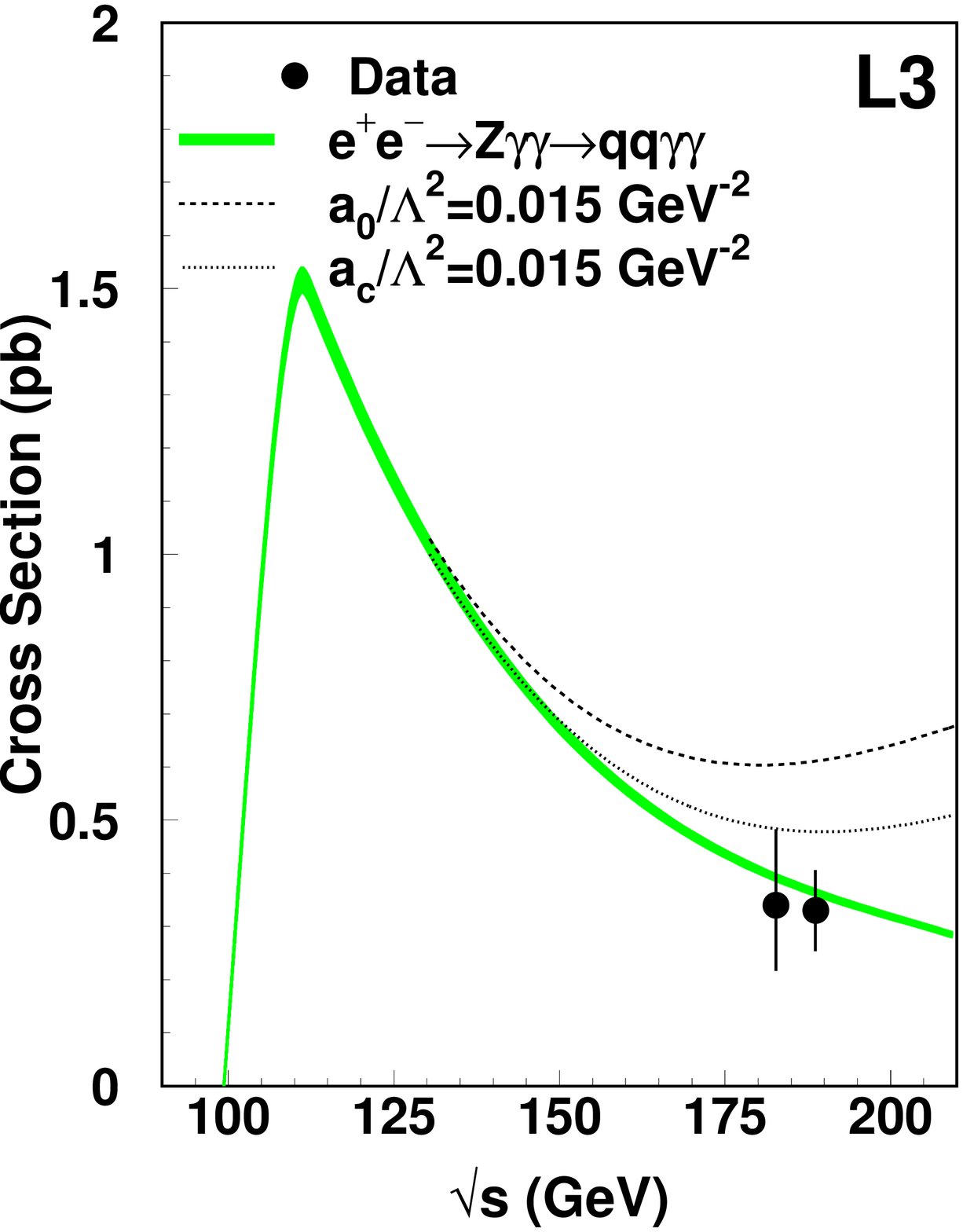}}
    \icaption{Evolution of the $\epem\ra\Zo\gamma\gamma\ra\qqbar\gamma\gamma$ 
      cross section with the centre-of-mass energy. Signal
    definition cuts described in the text are applied. The width of
    the band corresponds to the error that arises from MC statistics and
    theory uncertainty, estimated to be 1.5\%. Dashed and dotted lines
    represent QGC predictions.
    \label{fig:3}}
  \end{center}
\end{figure}

The cross
section  results are:

\begin{center}
\begin{tabular}{cccc}
$\sigma_{\epem\ra\Zo\gamma\gamma\ra\qqbar\gamma\gamma}(183\GeV)$ & = & $0.34^{+0.14}_{-0.12}\pm
0.03\,{\mathrm{pb}}$ & $(\rm SM\,\,\,0.396\pm 0.005\,{\mathrm{pb}})$\phantom{,}\\
$\sigma_{\epem\ra\Zo\gamma\gamma\ra\qqbar\gamma\gamma}(189\GeV)$ & =
& $0.33\pm 0.07\pm
0.03\,{\mathrm{pb}}$ & $(\rm SM\,\,\,0.365\pm 0.003\,{\mathrm{pb}})$,\\
\end{tabular}
\end{center}

\noindent
where the first uncertainties are statistical and the second systematic.
The values in parentheses denote the SM expectations calculated from the KK2f
MC with its default set of input parameters. The error on the
predictions is the quadratic sum of the MC statistical error and the
theory uncertainty estimated as suggested in Reference~\cite{KK2f}.
The measurements are in 
good agreement with these predictions. These results are also
presented in Figure~\ref{fig:3} together with the expected evolution
with $\sqrt{s}$ of the SM cross section.

Scaling the measured 
cross sections for the
Z hadronic branching ratio gives the $\epem\ra\Zo\gamma\gamma$ cross
section at 
$\sqrt{s}=183\GeV$ and $\sqrt{s}=189\GeV$:
\begin{center}
\begin{tabular}{ccc}
$\sigma_{\epem\ra\Zo\gamma\gamma}(183\GeV)$ & = & $0.49^{+0.20}_{-0.17}\pm
0.04\,{\mathrm{pb}}$\\
$\sigma_{\epem\ra\Zo\gamma\gamma}(189\GeV)$ & = & $0.47 \pm 0.10 \pm
0.04\,{\mathrm{pb}}$.\\
\end{tabular}
\end{center}

%
%%%%%%%%%%%%%%%%%%%%%%%%%%%%%%%%%%%%%%%%%%%%%%%%%%%%%%%%%%%%%%%%%%%%%%%%%%%%%%%
% Limits on Quartic Gauge Boson Coupling 
%%%%%%%%%%%%%%%%%%%%%%%%%%%%%%%%%%%%%%%%%%%%%%%%%%%%%%%%%%%%%%%%%%%%%%%%%%%%%%%
%
\section{Limits on Quartic Gauge Boson Couplings}

Anomalous QGC contributions to $\Zo\gamma\gamma$ production via the
$s$-channel exchange of a Z are described by two additional terms of
dimension six in the Lagrangian~\cite{bb1}:
\begin{eqnarray*}
{\cal L}^0_6 & = &  -{\pi\alpha \over 4\Lambda^2} a_0 F_{\mu\nu}F^{\mu\nu}
\vec{W}_\rho\cdot\vec{W}^\rho\\
{\cal L}^c_6 & = &  -{\pi\alpha \over 4\Lambda^2} a_c F_{\mu\rho}F^{\mu\sigma}
\vec{W}^\rho\cdot\vec{W}_\sigma,
\end{eqnarray*}
where $\alpha$ is the electromagnetic coupling, $F_{\mu\nu}$ is the
field strength tensor of the photon and $\vec{W}_\mu$ is the 
 weak boson
field. For $\Zo\gamma\gamma$ the third component of $\vec{W}_\mu$,
$Z_\mu/\cos{\theta_W}$, is relevant.
The parameters $a_0$ and $a_c$ describe the strength of the  QGC and
$\Lambda$ represents the scale of the New Physics responsible for the
coupling. ${\cal L}^0_6$ and 
${\cal L}^c_6$ are separately C and P conserving and no CP violating operators contribute to
the anomalous ZZ$\gamma\gamma$ vertex.
A more detailed description of QGC has recently appeared~\cite{bb2}.
While indirect limits on the QGC were derived from precision
measurements at the Z pole~\cite{eboli},
studies of $\Zo\gamma\gamma$ and $\Wp\Wm\gamma$ production probe the quantities 
$a_0/\Lambda^2$ and $a_c/\Lambda^2$ in a direct way. The
$\epem\ra\Zo\gamma\gamma$ 
process is expected to have higher sensitivities than
$\epem\ra\Wp\Wm\gamma$. This is due to an extra factor of $1/\cos^4{\theta_W}$
in the QGC cross section, to the larger SM
cross section and  data statistics  and to the smaller number of SM diagrams~\cite{sw}.

\begin{figure}[p]
  \begin{center}
    \begin{tabular}{c}
          \mbox{\includegraphics[width=.75\figwidth]{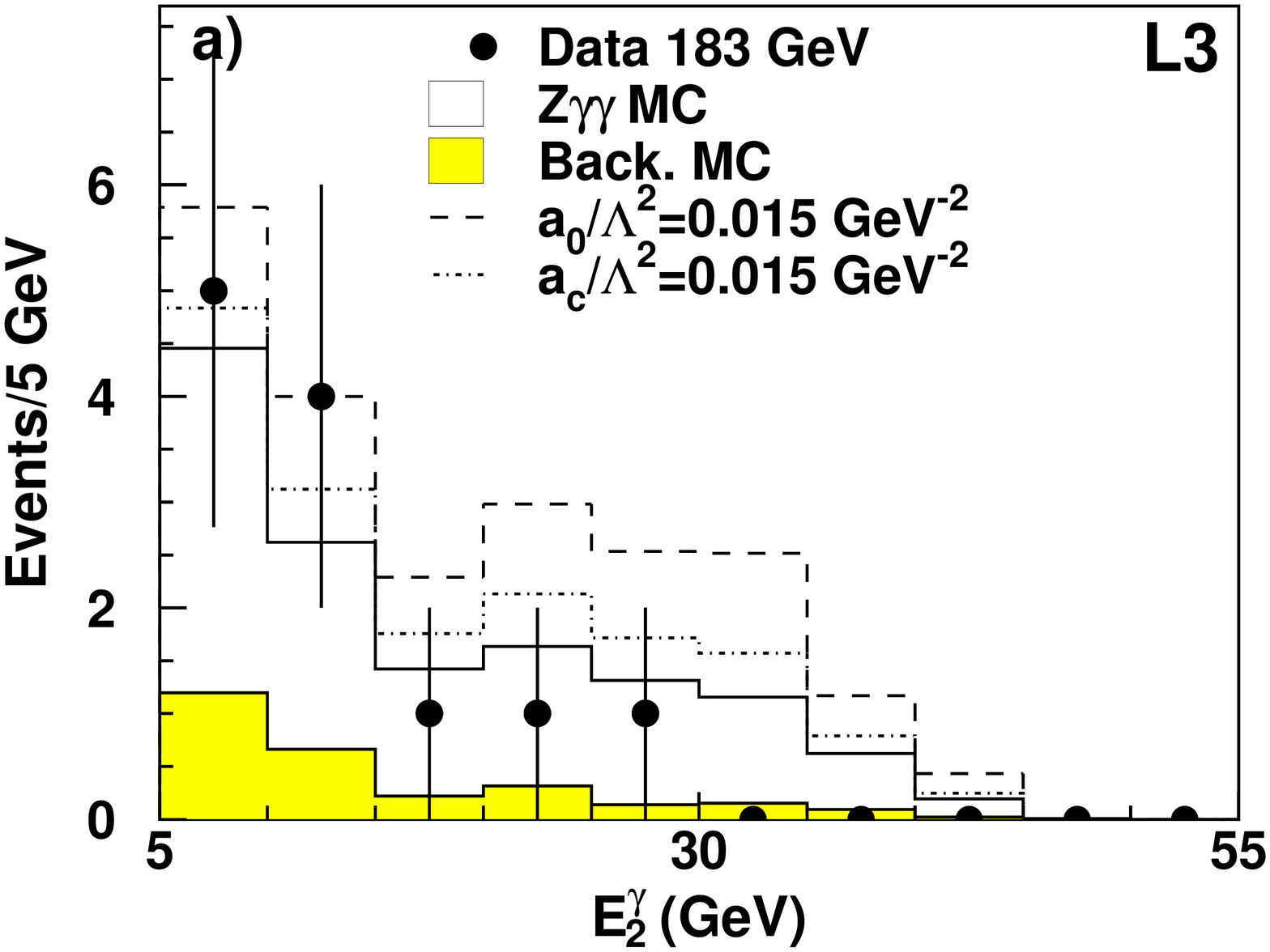}} \\
          \mbox{\includegraphics[width=.75\figwidth]{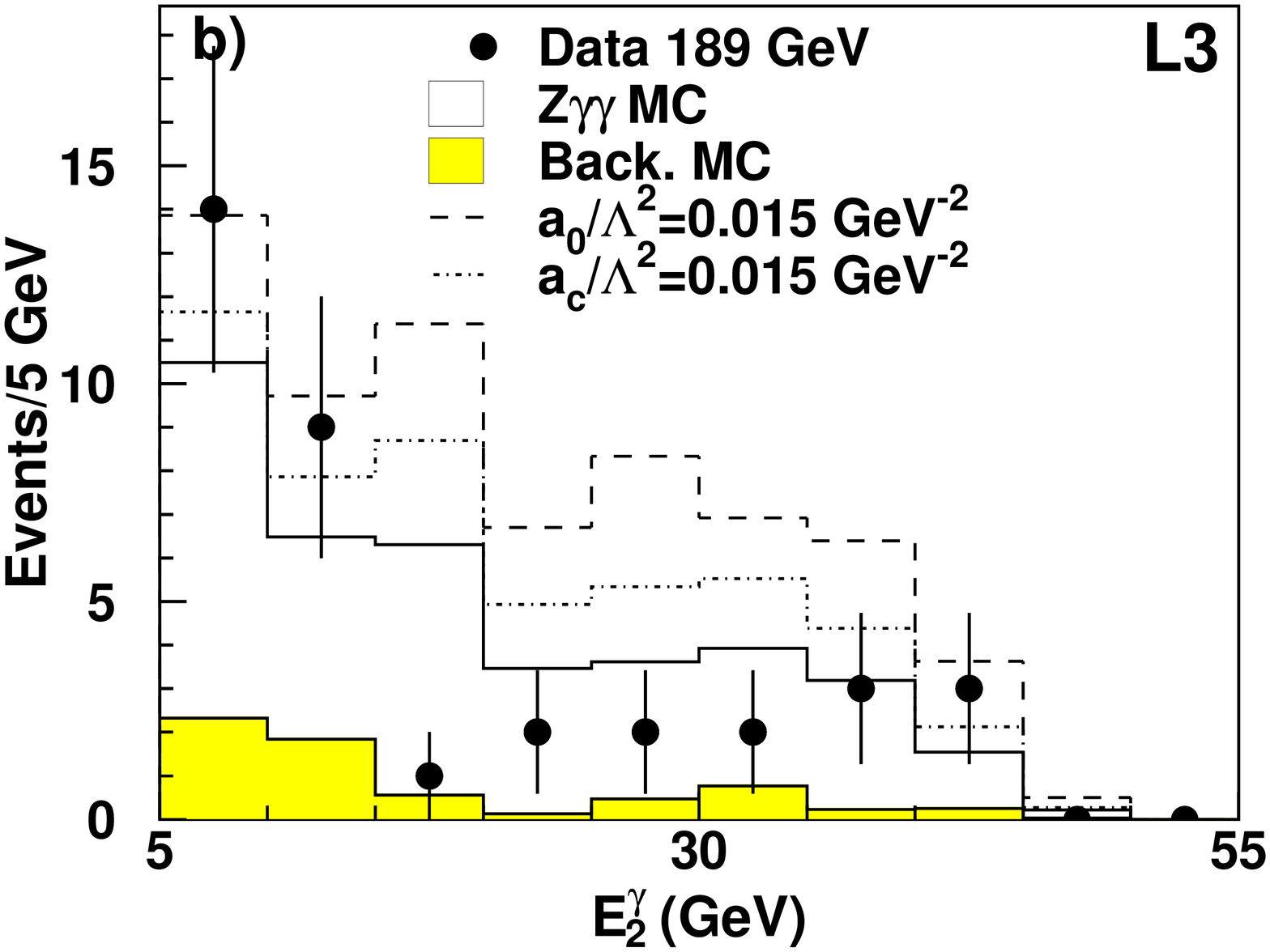}} \\
     \end{tabular}
    \icaption{Energy $E_2^\gamma$ of the least energetic photon for
    (a) $\sqrt{s} =183\GeV$ and (b) $\sqrt{s} =189\GeV$. Data,
    $\Zo\gamma\gamma$ and background MC 
    are displayed together with QGC predictions.
      \label{fig:4}}
  \end{center}
\end{figure}

QGC are expected to  manifest themselves via deviations in the total
$\epem\ra\Zo\gamma\gamma$ 
cross section, as  presented in Figure~\ref{fig:3}.
As the Z$\gamma\gamma$ production occurs in the SM via  $t$-channel
diagrams, the 
three body phase space favoured in the QGC mediated
production is different, in particular resulting in a  harder spectrum of the
least energetic photon~\cite{bb2}. Figures~\ref{fig:4}a and
\ref{fig:4}b compare these reconstructed spectra with the predictions
from signal and background MC at $\sqrt{s} = 183\GeV$ and  $\sqrt{s} =
189\GeV$. The  expectations for an anomalous value of $a_0/\Lambda^2$
or $a_c/\Lambda^2$ are also shown. 
These QGC predictions are obtained by reweighting each SM signal MC
event with the ratio ${\cal W}(\Omega,a_0/\Lambda^2,a_c/\Lambda^2)$,
a function of its  phase
space $\Omega$ derived from the two photons and the Z mass and the
values of the couplings:

\begin{displaymath}
{\cal W}(\Omega,a_0/\Lambda^2,a_c/\Lambda^2) = {|{\cal M}_{SM}(\Omega) + {\cal
  M}_{QGC}(\Omega,a_0/\Lambda^2,a_c/\Lambda^2) 
|^2\over |{\cal M}_{SM}(\Omega)|^2}.
\end{displaymath}
${\cal M}_{SM}$ denotes the SM matrix element and ${\cal M}_{QGC}$
the QGC one, both calculated analytically~\cite{sw}. Possible
extra initial state photons are taken into
account in the calculation of $\Omega$.

\begin{figure}[p]
  \begin{center}
      \mbox{\includegraphics[width=\figwidth]{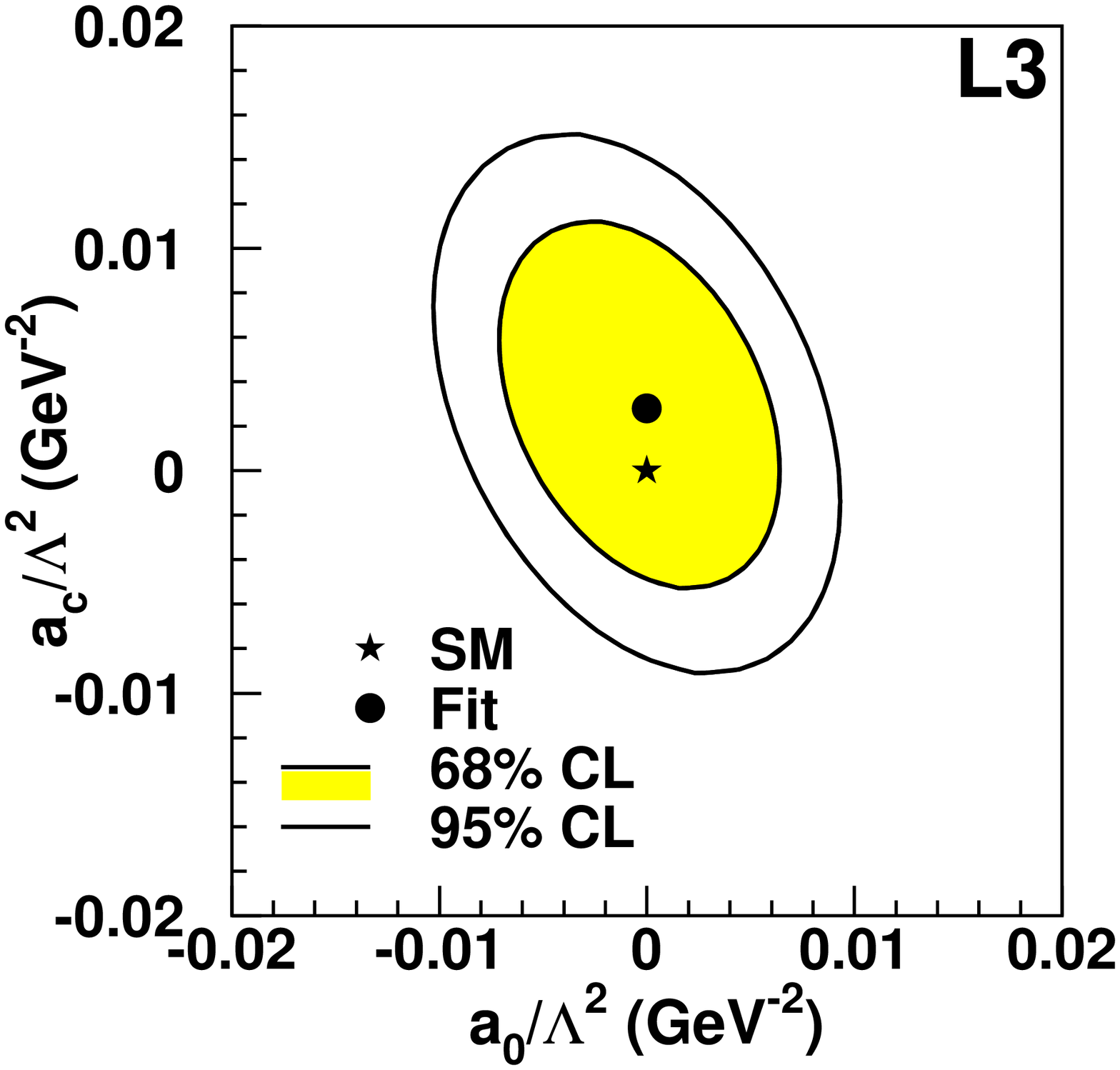}}
    \icaption{ Two dimensional
    contours for the QGC parameters $a_0/\Lambda^2$ and 
    $a_c/\Lambda^2$. 
      \label{fig:5}}
  \end{center}
\end{figure}

A simultaneous fit to the two energy spectra is performed leaving one of
the two QGC free at a time, fixing the other to zero. The SM
predictions in the fit procedure are reweighted as described above,
yielding the 68\% confidence level (CL) measurements:
\begin{displaymath}
a_0/\Lambda^2  =  0.001 \pm 0.004 \GeV^{-2}\,\,\,{\rm and}\,\,\,
a_c/\Lambda^2  =  0.003 \pm 0.005 \GeV^{-2}\,,
\end{displaymath}
in agreement with the expected SM value
of zero. 
A   simultaneous fit to both the parameters  yields
the 95\% CL limits:
\begin{displaymath}
-0.009\GeV^{-2} < a_0/\Lambda^2  < 0.008\GeV^{-2}\,\,\,{\rm and}\,\,\,
-0.007 \GeV^{-2}< a_c/\Lambda^2  < 0.013\GeV^{-2}\,,
\end{displaymath}
as shown in Figure~\ref{fig:5}.
A correlation of $-35\%$ is observed. The experimental systematic uncertainties and those on the SM
$\epem\ra\Zo\gamma\gamma\ra\qqbar\gamma\gamma$ 
cross section predictions are taken into account.

%
%%%%%%%%%%%%%%%%%%%%%%%%%%%%%%%%%%%%%%%%%%%%%%%%%%%%%%%%%%%%%%%%%%%%%%%%%%%%%%%
% Acknowledgements
%%%%%%%%%%%%%%%%%%%%%%%%%%%%%%%%%%%%%%%%%%%%%%%%%%%%%%%%%%%%%%%%%%%%%%%%%%%%%%%
%
\section*{Acknowledgements}

We are indebted to Anja Werthenbach for providing us with the
$\epem\ra\Zo\gamma\gamma$ analytical calculations  and for the inspiring discussions.
We wish to express our gratitude to the CERN accelerator divisions for the
superb performance and the continuous and successful upgrade of the
LEP machine.  
We acknowledge the contributions of the engineers  and technicians who
have participated in the construction and maintenance of this experiment.

%
%%%%%%%%%%%%%%%%%%%%%%%%%%%%%%%%%%%%%%%%%%%%%%%%%%%%%%%%%%%%%%%%%%%%%%%%%%%%%%%
% Bibliography
%%%%%%%%%%%%%%%%%%%%%%%%%%%%%%%%%%%%%%%%%%%%%%%%%%%%%%%%%%%%%%%%%%%%%%%%%%%%%%
%

%
%%%%%%%%%%%%%%%%%%%%%%%%%%%%%%%%%%%%%%%%%%%%%%%%%%%%%%%%%%%%%%%%%%%%%%%%%%%%%%
% Author List
%%%%%%%%%%%%%%%%%%%%%%%%%%%%%%%%%%%%%%%%%%%%%%%%%%%%%%%%%%%%%%%%%%%%%%%%%%%%%%
%

\newpage
\typeout{   }     
\typeout{Using author list for paper 199 ONLY!!!!!!!!!!!!!!!!!}
\typeout{Using author list for paper 199 ONLY!!!!!!!!!!!!!!!!!}
\typeout{Using author list for paper 199 ONLY!!!!!!!!!!!!!!!!!}
\typeout{Using author list for paper 199 ONLY!!!!!!!!!!!!!!!!!}
\typeout{Using author list for paper 199 ONLY!!!!!!!!!!!!!!!!!}
\typeout{Using author list for paper 199 ONLY!!!!!!!!!!!!!!!!!}
\typeout{$Modified: Wed Jan  5 17:33:47 2000 by clare $}
\typeout{!!!!  This should only be used with document option a4p!!!!}
\typeout{   }
%
%
%
%  L A T E X  version!!
%
%
% Make sure that the Lep package has been used!
%\input{Lep.sty}%
%
%\ifx\LepCalled\undefined%
%\typeout{     }%
%\typeout{!!!!!!!!!!!!!!!!!!!!!!!!!!!!!!!!!!!!!!!!!!!!!!!!!!!!!!!!!!!}%
%\typeout{Yikes.  You haven't used the Lep package!}%
%\typeout{Please put \protect\usepackage\protect{Lep\protect} in your preamble,
%         followed by}%
%\typeout{\protect\Lep\protect{1\protect} or \protect\Lep\protect{2\protect}}%
%\typeout{     }%
%\typeout{For now you will get a Lep phase 2 authorlist (may not be right!).}%
%\typeout{!!!!!!!!!!!!!!!!!!!!!!!!!!!!!!!!!!!!!!!!!!!!!!!!!!!!!!!!!!!}%
%\typeout{     }%
%\Lep{2}\fi%

\newcount\tutecount  \tutecount=0
\def\tutenum#1{\global\advance\tutecount by 1 \xdef#1{\the\tutecount}}
\def\tute#1{$^{#1}$}
\tutenum\aachen            % 1
\tutenum\nikhef            % 2
\tutenum\mich              % 3
\tutenum\lapp              % 4
\tutenum\basel             % 5
\tutenum\lsu               % 6
\tutenum\beijing           % 7
\tutenum\berlin            % 8
\tutenum\bologna           % 9 
\tutenum\tata              % 10
\tutenum\ne                % 11
\tutenum\bucharest         % 12
\tutenum\budapest          % 13
\tutenum\mit               % 14 
\tutenum\debrecen          % 15
\tutenum\florence          % 16
\tutenum\cern              % 17 
\tutenum\wl                % 18 
\tutenum\geneva            % 19
\tutenum\hefei             % 20
\tutenum\seft              % 21
\tutenum\lausanne          % 22
\tutenum\lecce             % 23
\tutenum\lyon              % 24
\tutenum\madrid            % 25
\tutenum\milan             % 26
\tutenum\moscow            % 27
\tutenum\naples            % 27
\tutenum\cyprus            % 29
\tutenum\nymegen           % 30
\tutenum\caltech           % 31
\tutenum\perugia           % 32
\tutenum\cmu               % 33
\tutenum\prince            % 34
\tutenum\rome              % 35
\tutenum\peters            % 36
\tutenum\potenza           % 37
\tutenum\salerno           % 38
\tutenum\ucsd              % 39
\tutenum\santiago          % 40
\tutenum\sofia             % 41 
\tutenum\korea             % 42
\tutenum\alabama           % 43
\tutenum\utrecht           % 44
\tutenum\purdue            % 45
\tutenum\psinst            % 46
\tutenum\zeuthen           % 47
\tutenum\eth               % 48
\tutenum\hamburg           % 49
\tutenum\taiwan            % 50
\tutenum\tsinghua          % 51
{
\parskip=0pt
\noindent
{\bf The L3 Collaboration:}
\ifx\selectfont\undefined%  old style font selection
 \baselineskip=10.8pt
 \baselineskip\baselinestretch\baselineskip
 \normalbaselineskip\baselineskip
 \ixpt
\else%                      new style font selection
 \fontsize{9}{10.8pt}\selectfont
\fi
\medskip
\tolerance=10000
\hbadness=5000
\raggedright
\hsize=162truemm\hoffset=0mm
\def\r{\rlap,}
\noindent

M.Acciarri\r\tute\milan\
P.Achard\r\tute\geneva\ 
O.Adriani\r\tute{\florence}\ 
M.Aguilar-Benitez\r\tute\madrid\ 
J.Alcaraz\r\tute\madrid\ 
G.Alemanni\r\tute\lausanne\
J.Allaby\r\tute\cern\
A.Aloisio\r\tute\naples\ 
M.G.Alviggi\r\tute\naples\
G.Ambrosi\r\tute\geneva\
% ...
\newpage
%\rule{\textwidth}{0.4pt}
\begin{list}{A}{\itemsep=0pt plus 0pt minus 0pt\parsep=0pt plus 0pt minus 0pt
                \topsep=0pt plus 0pt minus 0pt}
\item[\aachen]
 I. Physikalisches Institut, RWTH, D-52056 Aachen, FRG$^{\S}$\\
 III. Physikalisches Institut, RWTH, D-52056 Aachen, FRG$^{\S}$
\item[\nikhef] National Institute for High Energy Physics, NIKHEF, 
     and University of Amsterdam, NL-1009 DB Amsterdam, The Netherlands
\item[\mich] University of Michigan, Ann Arbor, MI 48109, USA
\item[\lapp] Laboratoire d'Annecy-le-Vieux de Physique des Particules, 
     LAPP,IN2P3-CNRS, BP 110, F-74941 Annecy-le-Vieux CEDEX, France
\item[\basel] Institute of Physics, University of Basel, CH-4056 Basel,
     Switzerland
\item[\lsu] Louisiana State University, Baton Rouge, LA 70803, USA
\item[\beijing] Institute of High Energy Physics, IHEP, 
  100039 Beijing, China$^{\triangle}$ 
\item[\berlin] Humboldt University, D-10099 Berlin, FRG$^{\S}$
\item[\bologna] University of Bologna and INFN-Sezione di Bologna, 
     I-40126 Bologna, Italy
\item[\tata] Tata Institute of Fundamental Research, Bombay 400 005, India
\item[\ne] Northeastern University, Boston, MA 02115, USA
\item[\bucharest] Institute of Atomic Physics and University of Bucharest,
     R-76900 Bucharest, Romania
\item[\budapest] Central Research Institute for Physics of the 
     Hungarian Academy of Sciences, H-1525 Budapest 114, Hungary$^{\ddag}$
\item[\mit] Massachusetts Institute of Technology, Cambridge, MA 02139, USA
\item[\debrecen] KLTE-ATOMKI, H-4010 Debrecen, Hungary$^\P$
\item[\florence] INFN Sezione di Firenze and University of Florence, 
     I-50125 Florence, Italy
\item[\cern] European Laboratory for Particle Physics, CERN, 
     CH-1211 Geneva 23, Switzerland
\item[\wl] World Laboratory, FBLJA  Project, CH-1211 Geneva 23, Switzerland
\item[\geneva] University of Geneva, CH-1211 Geneva 4, Switzerland
\item[\hefei] Chinese University of Science and Technology, USTC,
      Hefei, Anhui 230 029, China$^{\triangle}$
\item[\seft] SEFT, Research Institute for High Energy Physics, P.O. Box 9,
      SF-00014 Helsinki, Finland
\item[\lausanne] University of Lausanne, CH-1015 Lausanne, Switzerland
\item[\lecce] INFN-Sezione di Lecce and Universit\'a Degli Studi di Lecce,
     I-73100 Lecce, Italy
\item[\lyon] Institut de Physique Nucl\'eaire de Lyon, 
     IN2P3-CNRS,Universit\'e Claude Bernard, 
     F-69622 Villeurbanne, France
\item[\madrid] Centro de Investigaciones Energ{\'e}ticas, 
     Medioambientales y Tecnolog{\'\i}cas, CIEMAT, E-28040 Madrid,
     Spain${\flat}$ 
\item[\milan] INFN-Sezione di Milano, I-20133 Milan, Italy
\item[\moscow] Institute of Theoretical and Experimental Physics, ITEP, 
     Moscow, Russia
\item[\naples] INFN-Sezione di Napoli and University of Naples, 
     I-80125 Naples, Italy
\item[\cyprus] Department of Natural Sciences, University of Cyprus,
     Nicosia, Cyprus
\item[\nymegen] University of Nijmegen and NIKHEF, 
     NL-6525 ED Nijmegen, The Netherlands
\item[\caltech] California Institute of Technology, Pasadena, CA 91125, USA
\item[\perugia] INFN-Sezione di Perugia and Universit\'a Degli 
     Studi di Perugia, I-06100 Perugia, Italy   
\item[\cmu] Carnegie Mellon University, Pittsburgh, PA 15213, USA
\item[\prince] Princeton University, Princeton, NJ 08544, USA
\item[\rome] INFN-Sezione di Roma and University of Rome, ``La Sapienza",
     I-00185 Rome, Italy
\item[\peters] Nuclear Physics Institute, St. Petersburg, Russia
\item[\potenza] INFN-Sezione di Napoli and University of Potenza, 
     I-85100 Potenza, Italy
\item[\salerno] University and INFN, Salerno, I-84100 Salerno, Italy
\item[\ucsd] University of California, San Diego, CA 92093, USA
\item[\santiago] Dept. de Fisica de Particulas Elementales, Univ. de Santiago,
     E-15706 Santiago de Compostela, Spain
\item[\sofia] Bulgarian Academy of Sciences, Central Lab.~of 
     Mechatronics and Instrumentation, BU-1113 Sofia, Bulgaria
\item[\korea] Center for High Energy Physics, Adv.~Inst.~of Sciences
     and Technology, 305-701 Taejon,~Republic~of~{Korea}
\item[\alabama] University of Alabama, Tuscaloosa, AL 35486, USA
\item[\utrecht] Utrecht University and NIKHEF, NL-3584 CB Utrecht, 
     The Netherlands
\item[\purdue] Purdue University, West Lafayette, IN 47907, USA
\item[\psinst] Paul Scherrer Institut, PSI, CH-5232 Villigen, Switzerland
\item[\zeuthen] DESY, D-15738 Zeuthen, 
     FRG
\item[\eth] Eidgen\"ossische Technische Hochschule, ETH Z\"urich,
     CH-8093 Z\"urich, Switzerland
\item[\hamburg] University of Hamburg, D-22761 Hamburg, FRG
\item[\taiwan] National Central University, Chung-Li, Taiwan, China
\item[\tsinghua] Department of Physics, National Tsing Hua University,
      Taiwan, China
\item[\S]  Supported by the German Bundesministerium 
        f\"ur Bildung, Wissenschaft, Forschung und Technologie
\item[\ddag] Supported by the Hungarian OTKA fund under contract
numbers T019181, F023259 and T024011.
\item[\P] Also supported by the Hungarian OTKA fund under contract
  numbers T22238 and T026178.
\item[$\flat$] Supported also by the Comisi\'on Interministerial de Ciencia y 
        Tecnolog{\'\i}a.
\item[$\sharp$] Also supported by CONICET and Universidad Nacional de La Plata,
        CC 67, 1900 La Plata, Argentina.
\item[$\diamondsuit$] Also supported by Panjab University, Chandigarh-160014, 
        India.
\item[$\triangle$] Supported by the National Natural Science
  Foundation of China.
\item[\dag] Deceased.
\end{list}
}
\vfill

%%% Local Variables: 
%%% mode: latex
%%% TeX-master: t
%%% End:

%%% Local Variables: 
%%% mode: latex
%%% TeX-master: t
%%% TeX-master: t
%%% TeX-master: t
%%% End: 

%%% Local Variables: 
%%% mode: latex
%%% TeX-master: t
%%% End: 

%%% Local Variables: 
%%% mode: latex
%%% TeX-master: t
%%% End: 

%%% Local Variables: 
%%% mode: latex
%%% TeX-master: t
%%% End: 

\newpage

%
%%%%%%%%%%%%%%%%%%%%%%%%%%%%%%%%%%%%%%%%%%%%%%%%%%%%%%%%%%%%%%%%%%%%%%%%%%%%%%%
% Figures
%%%%%%%%%%%%%%%%%%%%%%%%%%%%%%%%%%%%%%%%%%%%%%%%%%%%%%%%%%%%%%%%%%%%%%%%%%%%%%%
%

%%%%%%%%%%%%%%%%%%%%%%%%%%%%%%%%%%%%%%%%%%%%%%%%%%%%%%%%%%%%%%%%%%%%%%%%%%%%%%%

\begin{thebibliography}{99}

\bibitem{sm_glashow}
S.~L. Glashow,
  Nucl. Phys. {\bf 22}  (1961) 579;
A. Salam,
  in Elementary Particle Theory, ed. {N.~Svartholm},  (Alm\-qvist and
  Wiksell, Stockholm, 1968), p. 367;
S. Weinberg,
  Phys. Rev. Lett. {\bf 19}  (1967) 1264.

\bibitem{opalwwg}
OPAL Collab., G.~Abbiendi \etal, Preprint CERN-EP/99-130, hep-ex/9910069;
  submitted to Phys. Lett. B.

\bibitem{bb1}
G.~B\'elanger and F.~Boudjema,
  Phys. Lett. {\bf B 288}  (1992) 201.

\bibitem{sw}
W.~J.~Stirling and A.~Werthenbach, Preprint DTP/99/30, hep-ph/9903315.

\bibitem{bb2}
G.~B\'elanger \etal, Preprint LAPTH-744/19, hep-ph/9908254.

\bibitem{l3_00}
L3 Collab., B.~Adeva \etal,
  Nucl. Inst. Meth. {\bf A 289}  (1990) 35;
L3 Collab., O.~Adriani \etal,
  Physics Reports {\bf 236}  (1993) 1;
I.~C.~Brock \etal,
  Nucl. Instr. and Meth. {\bf A 381}  (1996) 236;
M.~Chemarin \etal,
  Nucl. Inst. Meth. {\bf A 349}  (1994) 345;
M.~Acciarri \etal,
  Nucl. Inst. Meth. {\bf A 351}  (1994) 300;
A.~Adam \etal,
  Nucl. Inst. Meth. {\bf A 383}  (1996) 342;
G.~Basti \etal,
  Nucl. Inst. Meth. {\bf A 374}  (1996) 293.

\bibitem{KK2f}
S.~Jadach, B.F.L.~Ward and Z.~W\c{a}s, Preprint CERN-TH/99-235,
  hep-ph/9912214.

\bibitem{pythia}
T. Sj{\"o}strand, CERN--TH/7112/93 (1993), revised August 1995; T.
  Sj{\"o}strand, Comp. Phys. Comm. {\bf 82} (1994) 74.

\bibitem{koralz}
S.~Jadach, B.~F.~L.~Ward and Z.~W\c{a}s,
  Comp. Phys. Comm {\bf 79}  (1994) 503.

\bibitem{phojet}
R.~Engel, Z. Phys. {\bf C 66} (1995) 203; R.~Engel and J.~Ranft, Phys. Rev.
  {\bf D 54} (1996) 4244.

\bibitem{koralw}
M. Skrzypek \etal, Comp. Phys. Comm. {\bf 94} (1996) 216; M. Skrzypek \etal,
  Phys. Lett. {\bf B 372} (1996) 289.

\bibitem{exca}
R. Kleiss and R. Pittau, Comp. Phys. Comm. {\bf 85} (1995) 447; R. Pittau,
  Phys. Lett. {\bf B 335} (1994) 490.

\bibitem{geant}
The L3 detector simulation is based on GEANT Version 3.15. R. Brun \etal,
  ``GEANT 3'', CERN--DD/EE/84--1 (Revised), 1987. The GHEISHA program (H.
  Fesefeldt, RWTH Aachen Report PITHA 85/02 (1985)) is used to simulate
  hadronic interactions.

\bibitem{grace}
T.~Ishikawa \etal, KEK Report 92-19.

\bibitem{eboli}
A.~Brunstein, O.~J.~P. \'Eboli and M.~C.~Gonzales-Garcia,
  Phys. Lett. {\bf B 375}  (1996) 233.

\end{thebibliography}
\end{document}